\documentclass[preprints,article,accept,moreauthors,10pt,a4paper]{mdpi}
\usepackage{amsfonts,amssymb,amsmath,amsthm}
\usepackage{graphicx,epsfig}
\usepackage{epstopdf}
\usepackage[normalem]{ulem}
\usepackage{bbm,bm}
\usepackage[utf8]{inputenc}
\usepackage{hyperref}
\newcommand{\ket}[1]{\mbox{$ | #1 \rangle $}}
\newcommand{\bra}[1]{\mbox{$ \langle #1 | $}}

\newcommand{\tr}{\mathrm{tr}}
\newcommand{\Exp}[1]{\mathrm{e}^{\mbox{\footnotesize$#1$}}}
\newcommand{\I}{\mathrm{i}}
\newcommand{\Cr}{{\cal C}_r}
\newcommand{\Cl}{{\cal C}_{l_1}}

\firstpage{1}
\pubvolume{xx}
\issuenum{1}
\articlenumber{1}
\pubyear{2018}
\copyrightyear{2018}
\history{}
\Title{Coherence Depletion in Quantum Algorithms}

\Author{Ye-Chao Liu, Jiangwei Shang* and Xiangdong Zhang}

\AuthorNames{Ye-Chao Liu, Jiangwei Shang and Xiangdong Zhang}

\address[1]{%
Beijing Key Laboratory of Nanophotonics and Ultrafine Optoelectronic Systems, School of Physics, Beijing Institute of Technology, Beijing 100081, China}

\corres{\hangafter=1 \hangindent=1.05em \hspace{-0.82em} Correspondence: jiangwei.shang@bit.edu.cn}

\date{\today}
\abstract{Besides the superior efficiency compared to their classical counterparts, quantum algorithms known so far are basically task-dependent, and scarcely any common features are shared between them.
In this work, however, we show that the depletion of quantum coherence turns out to be a common phenomenon in these algorithms.
For all the quantum algorithms that we investigated including Grover's algorithm, Deutsch-Jozsa algorithm and Shor's algorithm, quantum coherence of the system states reduces to the minimum along with the successful execution of the respective processes.
Notably, a similar conclusion cannot be drawn using other quantitative measures such as quantum entanglement.
Thus, we expect that coherence depletion as a common feature can be useful for devising new quantum algorithms in the future.
}


\begin{document}

\section{Introduction}\label{sec:Intro}
The emergence of quantum algorithms that are able to solve problems exponentially faster than any classical algorithms is one of the leading incentives for the rapid development of quantum information science over the last three decades.
Especially exciting is the new concept of computing that makes use of quantum fundamental principles, coined quantum computing \cite{QCQI}.
In 1992, the Deutsch-Jozsa (DJ) algorithm \cite{Deutsch553_DJ_original} was first proposed, which can confirm a given function's type with only one single evaluation, compared to at worst $2^{n-1}+1$ ($n$ being the number of two-valued digits) queries by any possible classical algorithms.
Moreover, the DJ algorithm is deterministic in the sense that it can always produce the correct answer, which greatly improves the original solution by Deutsch \cite{Deutsch1985} that can only succeed with probability of one half.
Soon, the basic problem of factoring a large integer was offered a new quantum solution, that is, Shor's algorithm \cite{Shor1994}.
The exponentially faster speed-up over any classical approaches could be used to break public-key cryptography schemes such as the widely-used Rivest–Shamir–Adleman (RSA) scheme once a quantum computer were built.
Then, it is Grover's search algorithm \cite{Grover1996} which is used to locate a target item in an unsorted database.
For this problem, Grover's algorithm runs only quadratically faster compared to any classical algorithms, but it has been proven to be asymptotically optimal \cite{Grover2}.

Coincidentally, all the quantum algorithms mentioned above were proposed in the 1990s.
Since the dawn of this century, however, few new speed-up quantum algorithms have been designed that are comparable in impact with the existing ones.
For two exceptional developments, see the DQC1 algorithm \cite{DQC1} and the HHL algorithm \cite{HHL} which we skip to consider in the current work.
One of the possible reasons lies in the fact that all the quantum algorithms known so far are basically task-dependent, in other words, they share very few common features, if there were any.
Along with this line, the series of works by Latorre and coauthors \cite{majorization1, majorization2, majorization3} reported that all known efficient quantum algorithms obey a majorization principle (See Ref.~\cite{Flamini2018observation} for a recent experimental realization of majorization.).
To be more precise, the time arrow in these algorithms is a majorization arrow, which is conjectured to be a sort of driving force for the respective processes.
Besides this one, there are no other general features being reported ever since.

In this paper, however, we present a new common feature underling the efficient quantum algorithms in terms of quantum coherence (see Sec.~\ref{sec:resource} for a brief review).
Note that we only consider the ideal scenario of no decoherence from the environment.
Specifically, we find that coherence of the system states all reduces to the minimum along with the successful execution of the respective algorithms.
In a rough sense, this is a ``coherence arrow'' in quantum algorithms, but with many flexibilities.
This feature is similar to the majorization principle, with the possible reason being that both the concepts of coherence and majorization
are basis dependent \cite{majorization_coherence1, majorization_coherence2}.
However, unlike the descriptive nature of majorization, quantum coherence can be computed quantitatively using various coherence measures.
In this aspect, the feature that we find with coherence is a more versatile tool compared to the majorization principle.
On the other hand, a similar conclusion cannot be drawn using other quantitative measures
including quantum entanglement, which may be argued that entanglement is basis independent \cite{majorization_entanglement}.
For instance, although works like Ref.~\cite{Jozsa1999quantum} showed that entanglement has some relations with the quantum advantage, others \cite{Jozsa03entanglement, Boyer17DQC1, Datta08DQC1} also demonstrated that the quantum speed-up can exist without entanglement.

Actually, the analysis of quantum algorithms using coherence is not new \cite{DJ_resource, Grover_resource1, Grover_resource2}, but the respective algorithms were considered independently in those works and a unified picture is missing.
For instance, in Ref.~\cite{DJ_resource} the author examined the role played by coherence as a resource in the Deutsch-Jozsa and related algorithms,
and found that the less of coherence there is, the worse the algorithm will perform.
Although from different perspectives, both of Refs.~\cite{Grover_resource1, Grover_resource2} reported that the success probability of Grover's algorithm relies on coherence.
Nevertheless, the results presented in this paper give a combined view of all the quantum algorithms known so far with coherence.

This paper is organized as follows.
In Sec.~\ref{sec:resource}, we review briefly the resource theory of quantum coherence, and introduce the commonly-used coherence measures.
Then we start with the investigation of Grover's algorithm in Sec.~\ref{sec:Grover}, where the evolution of quantum coherence is thoroughly analyzed.
Next, we move on to the Deutsch-Jozsa algorithm in Sec.~\ref{sec:DJ} and Shor's algorithm in Sec.~\ref{sec:Shor}.
In Sec.~\ref{sec:Dis}, the consequences of coherence played in quantum algorithms are discussed, along with a comparison with other quantitative measures such as quantum entanglement.
We close with a short conclusion in Sec.~\ref{sec:Con}.

\section{Resource theory of quantum coherence}\label{sec:resource}
Along with the rapid development of quantum information science, an alternative way of assessing quantum phenomena as resources has appeared.
Consequently, many tasks that are not previously possible within the realm of classical physics may be now exploited with the new approach.
This resource-driven viewpoint has motivated the development of a quantitative theory that captures the resource character of physical properties in a mathematically rigorous manner.
The formulation of such resource theories was initially pursued with the quantitative theory of entanglement \cite{GUHNE20091, entanglment_resource2}, but has since spread to encompass many other operational settings, including quantum coherence \cite{resource_theory1, resource_theory2, resource_theory3}; see Ref.~\cite{coherence_resource} for a recent review.

Resource theory provides a unified framework for studying resource quantification and manipulations under restricted operations that are deemed free.
For coherence, we are restricted to incoherent operations, so only incoherent states are free. Recall that a state is incoherent if it is diagonal in the reference basis.
Recently, it has been demonstrated that coherence can be converted to other quantum resources, such as entanglement and discord by certain operations \cite{PhysRevLett.115.020403, PhysRevLett.116.160407, PhysRevLett.117.020402}.
However, compared to entanglement and discord, evidences show that coherence may be a potentially more fundamental quantum resource \cite{PhysRevA.92.022112}.
To quantify coherence, a rigorous framework has been proposed by Baumgratz \emph{et al.} in Ref.~\cite{PhysRevLett.113.140401}.
In this work, we employ the two most commonly-used coherence measures, namely the relative entropy of coherence and the $l_1$-norm of coherence.

The relative entropy of coherence \cite{PhysRevLett.113.140401} is defined as
\begin{equation}
  \Cr(\rho)=S(\rho_{\rm diag})-S(\rho)\,,
\end{equation}
where $S(\rho)=-\tr(\rho\log_2\rho)$ is the von Neumann entropy and $\rho_{\rm diag}=\sum_i\rho_{ii}\ket{i}\bra{i}$ denotes the state obtained from $\rho$ by deleting all the off-diagonal elements.
For pure states, the von Neumann entropy is 0, so the relative entropy can be simplified to
\begin{equation}
  \Cr(\rho)=S(\rho_{\rm diag}),\quad\mbox{if}\,\,\tr(\rho^2)=1\,.
\end{equation}
The $l_1$-norm of coherence \cite{PhysRevLett.113.140401} is defined intuitively as
\begin{equation}
  \Cl(\rho)=\sum_{i\neq j}|\rho_{ij}|\,,
\end{equation}
which comes from the fact that coherence is tied to the off-diagonal elements of the states.
Recently, it is demonstrated by Zhu \emph{et al.} \cite{PhysRevA.97.022342} that the $l_1$-norm of coherence is the analog of negativity in entanglement theory and sum negativity in the resource theory of magic-state quantum computation.
It is worth mentioning that both the relative entropy and the $l_1$-norm are proper measures of quantum coherence.

\section{Grover's algorithm}\label{sec:Grover}
We start with Grover's algorithm \cite{Grover1996}, which is a quantum search algorithm that runs quadratically faster than any equivalent classical algorithms.
Given an unsorted database with $N$ items, this algorithm is able to find the target item using only $O(\sqrt{N})$ steps, compared to at least $O(N)$ steps required by any classical schemes.
Although not offering an exponential speedup, Grover's algorithm has been proven to be asymptotically optimal for the search problem \cite{Grover2}.
For convenience, we assume $N=2^n$ such that the $N$ entries in the database can be supplied by $n$ qubits.
Let $f(x)$ be a function that takes in index $x=0,1,\dots,N-1$, and outputs $f(x)=1$ if $x$ is a solution to the search problem, and $f(x)=0$ otherwise.

Grover's algorithm begins with the initialized equal superposition state,
\begin{equation}\label{eq:psi0}
  \ket{\psi^{(0)}}=\frac1{\sqrt{2^n}}\sum_{x=0}^{2^n-1}\ket{x}\,,
\end{equation}
which has the maximal coherence
\begin{subequations}
\begin{align}
\Cr^{(0)}&=S(I/2^n)=n\,,\\
\Cl^{(0)}&=\sum_{i\ne j}^{2^{n}-1}\left|\frac{1}{2^{n}}\right|=2^{n}-1\,.
\end{align}
\end{subequations}
Suppose there are exactly $M$ solutions in the database with $1\leq M\leq N$, we can reexpress $\ket{\psi^{(0)}}$ as
\begin{equation}
\ket{\psi^{(0)}}=\sqrt{\frac{N-M}{N}}\ket{\alpha}+\sqrt{\frac{M}{N}}\ket{\beta}\,,
\end{equation}
where $\ket{\alpha}$ represents the group of states that are not solutions to the search problem (marked by $x_n$ below), while $\ket{\beta}$ represents those that are solutions (marked by $x_s$). Explicitly, we have
\begin{align}
\ket{\alpha}&\equiv \frac{1}{\sqrt{N-M}}\sum_{x_{n}}|x_{n}\rangle\,,\\
\ket{\beta}&\equiv \frac{1}{\sqrt{M}}\sum_{x_{s}}|x_{s}\rangle\,.
\end{align}

\begin{figure}[t]
\includegraphics[width=.5\columnwidth]{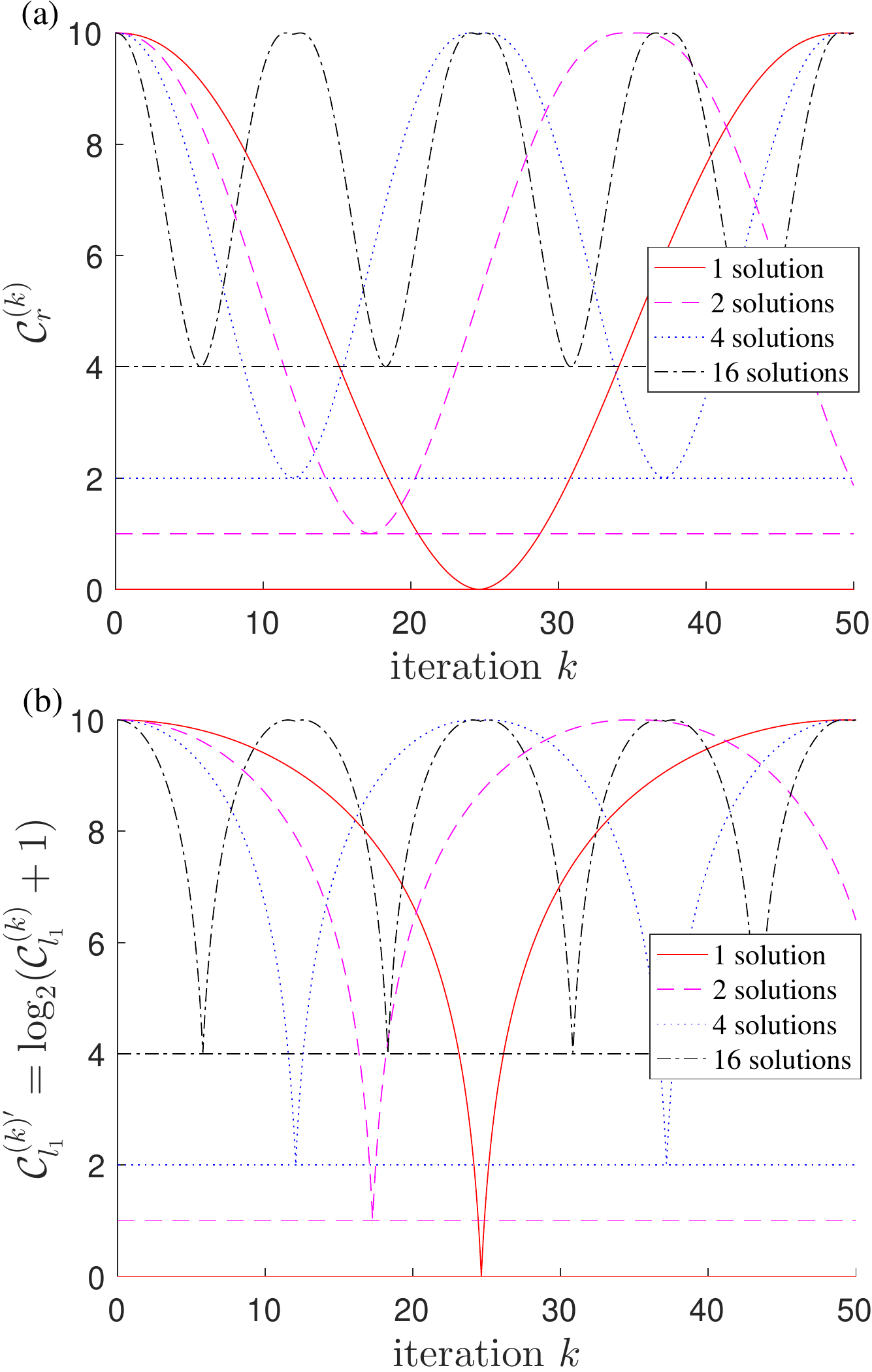}
\centering
\caption{\label{fig:GA_coherence}
In the case of ${n=10}$ qubits, we plot the values of coherence with respect to the number of Grover iterations $k$: (a) the relative entropy of coherence $\Cr^{(k)}$ in Eq.~\eqref{eq:Crk}; (b) $\Cl^{(k)'}=\log_2(\Cl^{(k)}+1)$ with $\Cl^{(k)}$ being the $l_1$-norm of coherence in Eq.~\eqref{eq:Clk}.
The plots show the results with $M=1,2,4$ and $16$ solutions respectively.
The minimal values indicate that an solution is found.
As we can see, with the number of possible solutions increased, not only the number of Grover iterations needed decreases, but also the minimal value of coherence increases accordingly.
See text and Fig.~\ref{fig:GA_reduction} for more details.
}
\end{figure}

Then, a subroutine known as the Grover iteration is applied to $\ket{\psi^{(0)}}$ repeatedly.
The Grover iteration consists of two basic operations $G=DO$, i.e.,
\begin{align}\label{eq:oracle}
O&:|x\rangle\to (-1)^{f(x)}\ket{x}\,,\\
D&=2\ket{\psi^{(0)}}\bra{\psi^{(0)}}-I\,,
\end{align}
where $O$ is an oracle (a black-box operation), and $D$ is the Grover diffusion operator.
After $k$ iterations of applying $G$, the state becomes
\begin{equation}
\ket{\psi^{(k)}}\equiv G^{k}\ket{\psi}=\cos\omega\ket{\alpha}+\sin\omega\ket{\beta}\,,
\end{equation}
where
\begin{equation}
\omega=\frac{2k+1}{2}\theta,\quad\mbox{with}\,\,\cos\!\left(\frac{\theta}{2}\right)=\sqrt{\frac{N-M}{N}}\,.
\end{equation}
It is not difficult to see that, with high probabilities, a solution to the search problem can be obtained by having $k=\left \lfloor \frac{\pi}{4} \sqrt{\frac{N}{M}}\right \rceil$ Grover iterations, where the symbol $\lfloor \cdot\rceil$ denotes the closest integer to the rational number inside.
Next, we calculate the quantum coherence of state \ket{\psi^{(k)}}, such that
\begin{subequations}\label{eq:Ck}
\begin{align}
\Cr^{(k)}&=-2\left(\cos^{2}\omega\log\frac{|\cos\omega|}{N-M}+\sin^{2}\omega\log\frac{|\sin\omega|}{M}\right),\label{eq:Crk}\\
\Cl^{(k)}&=\left(\sqrt{N-M}|\cos\omega|+\sqrt{M}|\sin\omega|\right)^{2}-1\,.\label{eq:Clk}
\end{align}
\end{subequations}
Note that the oracle $O$ only marks the solution by changing the phase of state $\ket{\beta}$, i.e.,
\begin{equation}
O\ket{\psi^{(k)}}=\cos\omega\ket{\alpha}-\sin\omega\ket{\beta}\,,
\end{equation}
so this operation will not change the coherence. It is the operation $D$ that indeed changes the coherence.

In Fig.~\ref{fig:GA_coherence}, we plot the values of coherence with respect to the number of Grover iterations $k$, for the case of $n=10$ qubits.
Note that in Fig.~\ref{fig:GA_coherence}(b), we plot instead $\Cl^{(k)'}=\log_2(\Cl^{(k)}+1)$.
As can be seen, a solution to the search problem is found when the coherence first reaches the minimum value, that is, when $k^{*}=\left \lfloor \frac{\pi}{4}\sqrt{\frac{N}{M}}\right \rceil$.
At this point, the task of Grover's algorithm is actually completed.
However, if the Grover iteration is continued, a periodic feature of the coherence appears such that we will get the solution again around $2k^{*}$ iterations.
This observation is repeated as long as the Grover iteration goes on.

\begin{figure}[t]
\centering
\includegraphics[width=.5\columnwidth]{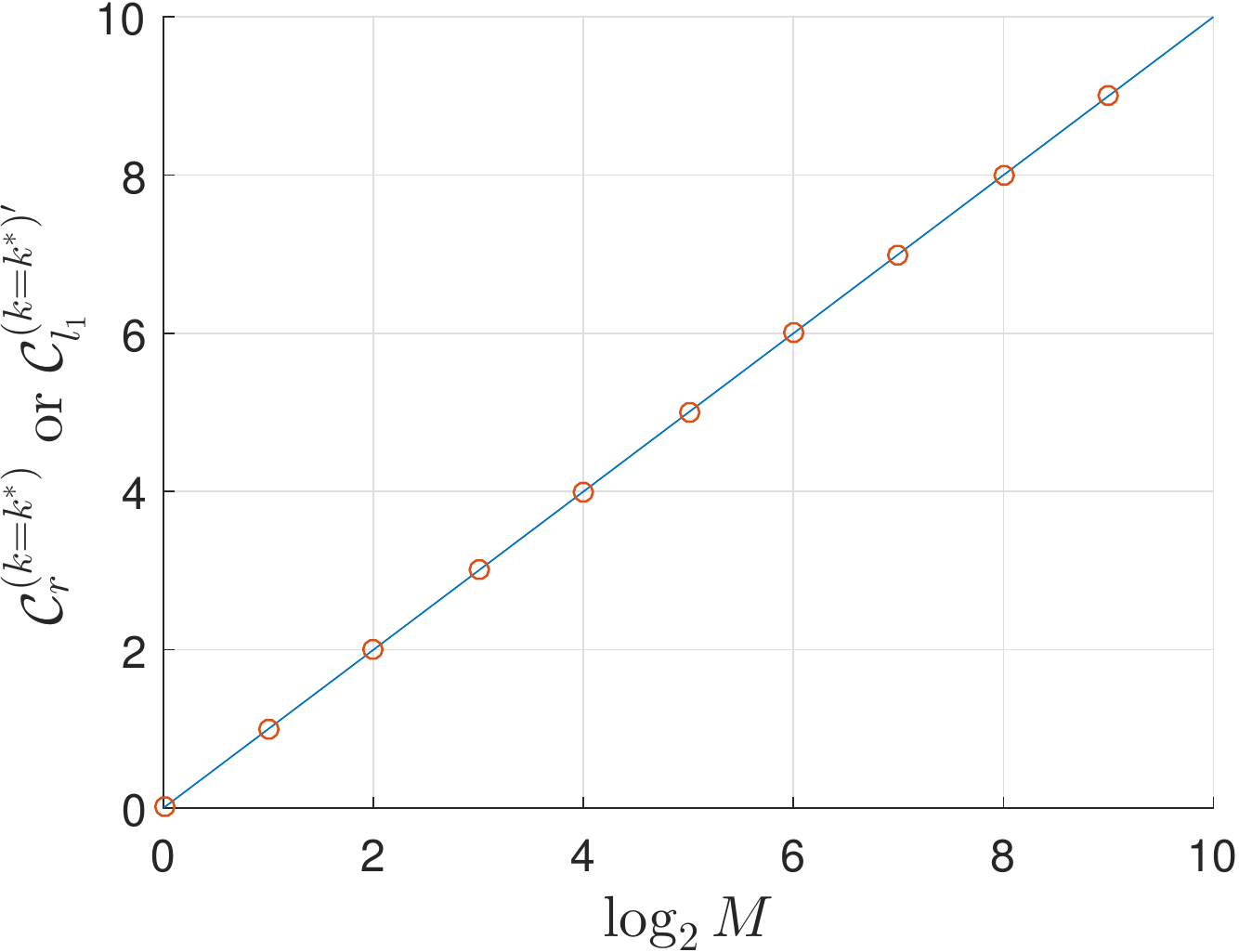}
\caption{\label{fig:GA_reduction}
Minimal coherence of the system state with respect to the logarithm of the number of solutions $\log_2 M$.
As we can see, with the number of solutions $M$ increased, the minimal value of coherence gets bigger which clearly indicates a superposition state consisting of more terms.
}
\end{figure}

Another phenomenon from the plots is as follows: with the number of possible solutions increased, not only the number of Grover iterations needed decreases, but also the minimal value of coherence gets bigger accordingly.
This is easy to understand as several answers ($M>1$) make up a superposition state of which the coherence is finite; see Fig.~\ref{fig:GA_reduction}.
To understand it better, let's look at the derivatives of the coherence in Eq.~\eqref{eq:Ck}, which are given by
\begin{subequations}
\begin{equation}
\frac{\mathrm{d}\Cr^{(k)}}{\mathrm{d}k}
=\theta\sin[(2k+1)\theta]\log\left[\frac{M}{N-M}\cot^{2}\left(\frac{2k+1}{2}\theta\right)\right],
\end{equation}
\begin{align}
\qquad\frac{\mathrm{d}\Cl^{(k)}}{\mathrm{d}k}&=\theta\Bigl\{(2M-N)\sin[(2k+1)\theta]\\
&+2\sqrt{(N-M)M}\,\mbox{sgn}\{\sin[(2k+1)\theta]\}\cos[(2k+1)\theta]\Bigr\}\,.\nonumber
\end{align}
\end{subequations}
By forcing the derivatives to be zero, we get four different cases:
\begin{enumerate}
  \item $\cos\left(\frac{2k+1}{2}\theta\right)=0$ corresponds to the minimal values in Fig.~\ref{fig:GA_coherence}, namely the solution state $\ket{\beta}$.
  \item $\cot^{2}\left(\frac{2k+1}{2}\theta\right)=\cot^{2}\left(\frac{\theta}{2}\right)$ corresponds to the maximal values in Fig.~\ref{fig:GA_coherence}. Because of the square in this solution, there are actually two peaks close to each other (not quite visible if the number of solutions $M$ is small). The right peak corresponds to the superposition state $\ket{\psi^{(0)}}$, while the left one corresponds to $O\ket{\psi^{(0)}}$.
  \item $\cos\left(\frac{2k+1}{2}\theta\right)=\pm1$ corresponds to the local minimal values between the two peaks in Fig.~\ref{fig:GA_coherence}.
      Because the distance between these two peaks is exactly 1 and we are considering discrete operations, so this local valley has no physical meaning.
  \item $\theta=0$ means that there is no solution, i.e., $M=0$.
\end{enumerate}

To summarize, one learns that the depletion of quantum coherence can be regarded
as a signal
for the successful executions of Grover's algorithm.
We will see later that the same conclusion can be drawn for other quantum algorithms including the Deutsch-Jozsa algorithm and Shor's algorithm.

\section{Deutsch–Jozsa algorithm}\label{sec:DJ}
Given a function $f(x)$ defined over the variable $x=0,1,\dots,2^n-1$ with $n$ being the number of dichotomic-valued digits, the Deutsch-Jozsa (DJ) algorithm \cite{Deutsch553_DJ_original} aims to confirm whether $f(x)$ is \emph{constant} for all values of $x$, or else it is \emph{balanced}, namely $f(x)=1$ for exactly half of all possible $x$, and 0 for the other half.
Although of little practical use, the DJ algorithm is deterministic in the sense that it can always produce the correct answer using only one correspondence, whereas it requires at worst $2^{n-1}+1$ queries for any possible classical algorithms.

Same as Grover's algorithm, the DJ algorithm begins by first preparing the equal superposition state of Eq.~\eqref{eq:psi0},
\begin{equation}\label{eq:psi0-DJ}
  \ket{\psi^{(0)}}=\frac1{\sqrt{2^n}}\sum_{x=0}^{2^n-1}\ket{x}\,,
\end{equation}
which has the maximal coherence
\begin{subequations}\label{eq:max_DJ}
\begin{align}
\Cr^{(0)}&=n\,,\\
\Cl^{(0)}&=2^{n}-1\,.
\end{align}
\end{subequations}
However, unlike Grover's algorithm, no iteration is needed in the DJ algorithm.
The next step is an oracle $U_f:\ket{x}\rightarrow(-1)^{f(x)}\ket{x}$ that transforms the state to
\begin{equation}
\ket{\psi^{(1)}}=\frac1{\sqrt{2^n}}\sum_{x=0}^{2^n-1}(-1)^{f(x)}\ket{x}\,,
\end{equation}
which leaves the coherence unchanged.
The final step of the DJ algorithm is to apply the Hadamard gate $H$, such that the state becomes
\begin{equation}
\ket{\psi^{(2)}}=\frac1{2^n}\sum_{y=0}^{2^n-1}\sum_{x=0}^{2^n-1}(-1)^{x\cdot y+f(x)}\ket{y}\,,
\end{equation}
where $x\cdot y$ is the bitwise inner product of $x$ and $y$.
Now, by examining the probability of measuring $\ket{0}^{\otimes n}$, i.e., $\left|\sum_x(-1)^{f(x)}/2^n\right|^2$, one gets 1 if $f(x)$ is constant and 0 if it is balanced.
Depending on the function type of $f(x)$, the coherence of $\ket{\psi^{(2)}}$ can have the following two cases:
\begin{enumerate}
\item If $f(x)$ is a constant function, then
\begin{equation}
\ket{\psi^{(2)}}=\frac1{2^n}
\left(
\begin{array}{cccc}
 \pm 2^n & 0  & \cdots & 0
\end{array}
\right)^T,
\end{equation}
the coherence of which is $\Cr^{(2)}=\Cl^{(2)}=0$.
\item If $f(x)$ is a balanced function, then
\begin{subequations}
\begin{align}
\Cr^{(2)} &\in [0,n-1]\,,\\
\Cl^{(2)} &\in [0,2^{n-1}-1]\,.
\end{align}
\end{subequations}
\end{enumerate}
For the second case, the coherence has a range instead of a single value due to the possible different forms of the balanced function.
For instance, if $f(x)$ takes values $01010101\cdots$ (for more than three qubits), then $\ket{\psi^{(2)}}$ is nothing but a basis state  with coherence being zero.
But, if $f(x)$ takes the sequence such as $01100101\cdots$, then \ket{\psi^{(2)}} is a superposition of basis states with nonzero coherence.
Notably, the coherence cannot take the maximal value as that in Eq.~\eqref{eq:max_DJ}, because the basis state $\ket{0}$ disappears in $\ket{\psi^{(2)}}$ for the balanced case.

Therefore, no matter what the function type of $f(x)$ is, we find that the coherence of the system state always decreases once the algorithm stops.
Again, coherence depletion can be used as a good signature to signal the success of the DJ algorithm.

\section{Shor's algorithm/Quantum order-finding}\label{sec:Shor}
Shor's algorithm \cite{Shor1994} is a particular instance of the family of quantum phase-estimation algorithms \cite{Cleve339}.
Informally, Shor's algorithm  solves the following problem: given an integer $N$, find its prime factors.
The crucial step in Shor's algorithm is the so-called quantum order-finding (QOF) subroutine which offers the quantum speedup over any classical approaches.
For two positive integers $x$ and $N$, the objective of QOF is to determine the \emph{order} of $x$ modulo $N$, which is defined as the least integer $r>0$, such that $x^r=1(\bmod \,N)$.

The QOF subroutine begins with $t=2L+1+\left\lceil\log(2+\frac1{2\epsilon})\right\rceil$ qubits initialized to $\ket{0}$ (the first register) and $L$ qubits initialized to $\ket{1}$ (the second register), where $L\equiv\lceil\log(N)\rceil$ denotes the closest integer larger than $\log(N)$ and $\epsilon$ is the error tolerance.
Application of the Hadamard gate $H$ on the first register transforms the initial state to
\begin{equation}
  \ket{\psi^{(0)}}=\frac1{2^t}\sum_{j=0}^{2^t-1}\ket{j}\ket{1}\,,
\end{equation}
which has the maximal coherence on the first $t$ qubits, i.e.,
\begin{subequations}
\begin{align}
\Cr^{(0)}&=t\,,\\
\Cl^{(0)}&=2^{t}-1\,.
\end{align}
\end{subequations}
Then a black box operation $U_{x,N}:\ket{j}\ket{k}\rightarrow\ket{j}\ket{x^j(\bmod \,N)}$ transforms the state to
\begin{equation}
  \ket{\psi^{(1)}}=\frac1{2^n}\sum_{j=0}^{2^t-1}\ket{j}\ket{x^j(\bmod \,N)}\,.
\end{equation}
Although the state $\ket{\psi^{(1)}}$ looks rather different from $\ket{\psi^{(0)}}$, its coherence (on the first $t$ qubits) does not change, namely
\begin{subequations}
\begin{align}
\Cr^{(1)}&=\Cr^{(0)}=t\,,\\
\Cl^{(1)}&=\Cl^{(0)}=2^{t}-1\,.
\end{align}
\end{subequations}
Because of the periodic nature of the component $\ket{x^j(\bmod \,N)}$, the state $\ket{\psi^{(1)}}$ can be approximated as
\begin{equation}
\ket{\psi^{(1)}}\approx\frac1{\sqrt{r2^n}}\sum_{s=0}^{r-1}\sum_{j=0}^{2^t-1}\Exp{2\pi\I\cdot sj/r}\ket{j}\ket{u_s}\,.
\end{equation}
The period of the phase in $\ket{\psi^{(1)}}$ can be obtained by applying inverse Fourier transform to the first register, such that
\begin{equation}
\ket{\psi^{(2)}}=\frac1{\sqrt{r}}\sum_{s=0}^{r-1}\ket{\widetilde{s/r}}\ket{u_{s}}\,,
\end{equation}
where $\ket{\widetilde{s/r}}$ is a pretty good approximation of the phase $s/r$.
Now, coherence of the state $\ket{\psi^{(2)}}$ becomes
\begin{subequations}
\begin{align}
\Cr^{(2)}&=-r^{2}\cdot\frac{1}{r^{2}}\log\frac{1}{r^{2}}=2\log r\,,\\
\Cl^{(2)}&=\frac{r^{4}-r^{2}}{r^{2}}=r^{2}-1\,,
\end{align}
\end{subequations}
which are functions of the solution $r$.
Finally, by measuring the first $t$ qubits, the solution $r$ is obtained by applying the continued fractions algorithm \cite{QCQI}.
Once again, we find that coherence of the system state reduces to the minimum by the end of the QOF subroutine, in turn also in Shor's algorithm.

\section{Discussion}\label{sec:Dis}
For all the quantum algorithms that we have explored including Grover's algorithm, DJ algorithm and Shor's algorithm, we find that quantum coherence plays a consistent role for signaling the completion of all processes.
To be more precise, upon successful executions of these algorithms, coherence of the respective systems all reduces to the minimum compared to the initial values.
Specifically, all the three quantum algorithms begin with the equal superposition state which has the maximal coherence.
Then, an oracle is applied, which leaves the coherence unchanged.
The final step can be seen as an adjustment for the system states, that is, diffusion operation for Grover's algorithm, Hadamard operation for DJ algorithm, and quantum inverse Fourier transform for Shor's algorithm.
It is this final step of operation that indeed reduces the coherence.
Hence, as a guide for future quantum-algorithm design, the coherence-depletion operation might be taken as an indispensable requirement for the relevant processes.

Then, it is natural to ask whether other quantitative measures such as entanglement may play a similar role as coherence in quantum algorithms.
Unfortunately, the answer is negative.
Many previous works have shown that a general principle cannot be drawn using entanglement.
For instance, in Refs.~\cite{Shor_ent1, Shor_ent2, Shor_ent3} the authors analyzed thoroughly the entanglement properties in Shor's algorithm, and found that entanglement may vary with different entanglement measures.
Then, similar conclusions were reported in Refs.~\cite{DJ_ent1, DJ_ent2, DJ_ent3, DJ_ent4} for the DJ algorithm.
In particular, Ref.~\cite{Jozsa03entanglement} showed that quantum algorithms can be efficiently simulated classically even when entanglement exists, whereas Refs.~\cite{Jozsa03entanglement, Boyer17DQC1, Datta08DQC1} demonstrated that quantum algorithms can show advantage without entanglement.
One of the possible reasons for the failure of using entanglement as a signature is due to the differences in definitions of entanglement and coherence (also majorization),
since both the concepts of coherence and majorization are basis dependent \cite{majorization_coherence1, majorization_coherence2}, while entanglement not.

No doubt that entanglement is a key (but not really sufficient \cite{entanglment_resource2}) resource for the quantum speedup in all these algorithms \cite{Jozsa1999quantum}, it cannot be used as a good signature in quantum algorithms.
Actually, it is an NP-hard problem \cite{NP1, NP2} to even detect entanglement if the system size is large, let alone to quantify it.
For multipartite quantum systems, entanglement can be classified into different forms depending on how the subsystems are distributed.
This complexity further makes any possible entanglement measures hard to compute.
Therefore, although essential for the quantum speedup, entanglement is not a good signature to use for quantum algorithms as compared to coherence.
Moreover, this fact also serves as an additional evidence that coherence may be a potentially more fundamental quantum resource than entanglement and discord, as initially argued in Ref.~\cite{PhysRevA.92.022112}.

\section{Conclusion}\label{sec:Con}
The scarceness of efficient quantum algorithms suggests that maybe some basic principles are missing.
In this paper, we find that the depletion of quantum coherence turns out to be a common feature in these algorithms.
For all the three quantum algorithms that we investigated including Grover's algorithm, Deutsch-Jozsa algorithm and Shor's algorithm, quantum coherence of the system states all reduces to the minimum along with the successful execution of the respective algorithms.
However, a similar conclusion cannot be drawn using other quantitative measures such as quantum entanglement.
Therefore, besides the fundamental interests in resource theory, this special feature of quantum coherence is expected to be useful for devising new quantum algorithms in future.


\vspace{6pt}

\authorcontributions{YCL performed the numerical calculations.
All authors contributed to the interpretation of the results, and writing of the manuscript.}

\funding{This work has been supported by the National Key R\&D Program of China under Grant No.~2017YFA0303800 and the National Natural Science Foundation of China through Grant Nos.~11574031 and 11805010.
J.S. also acknowledges support by the Beijing Institute of Technology Research Fund Program for Young Scholars.}

\acknowledgments{We thank Huangjun Zhu and Hui Khoon Ng for fruitful discussions. We are also grateful to Miguel Mart{\'i}n-Delgado for pointing out their recent work on majorization \cite{Flamini2018observation} to us.}

\conflictsofinterest{The authors declare no conflicts of interest.}

\reftitle{References}
\bibliographystyle{naturemag}
\bibliography{my_refs}

\end{document}